\documentstyle[aps]{revtex}
\input{epsf}

\begin{document}
\title{A model for the atomic-scale structure of a dense, nonequilibrium fluid:\
the homogeneous cooling state of granular fluids}
\author{James F. Lutsko}
\address{Center for Nonlinear Phenomena and Complex Systems\\
Universit\'{e} Libre de Bruxelles\\
1050-Bruxelles, Belgium\\
email:\ jim.lutsko@skynet.be}
\date{\today}
\maketitle

\begin{abstract}
It is shown that the equilibrium Generalized Mean Spherical Model of fluid
structure may be extended to nonequilibrium states with equation of state
information used in equilibrium replaced by an exact condition on the
two-body distribution function. The model is applied to the homogeneous
cooling state of granular fluids and upon comparison to molecular dynamics
simulations is found to provide an accurate picture of the pair distribution
function.
\end{abstract}


\section{Introduction}

The atomic scale structure, e.g. the pair distribution function or,
equivalently, the density-density equal time correlation function, of both
equilibrium and nonequilibrium fluids is directly accessible to experiment
by means of light scattering and has been used to study the behavior of
complex systems such as sheared colloidal suspensions\cite{Ackerson88}.
However, the subject of nonequilibrium fluid structure remains obscure,
particularly when compared to the structure of equilibrium fluids which is
one of the most advanced areas of equilibrium statistical mechanics\cite
{HansenMcdonald}. While kinetic theory can provide some asymptotic results,
see e.g. \cite{TK_LightScattering},\cite{Lutsko_ShearFluctuations},\cite
{Mirim}, models of the small-scale structure remain phenomenological. This
is not surprising: in equilibrium, one starts with the exact N-body
distribution and the problem is to integrate out uninteresting degrees of
freedom so as to get the pair distribution function (pdf). Away from
equilibrium, even in steady states, the N-body distribution function is not
known and the only starting point is either a complex dynamical equation,
the Liouville equation, or equivalently a coupled set of equations, the
BBGKY\ hierarchy relating the n-body to the (n+1)-body distribution
functions. Another method used to derive equilibrium integral equations is
based on the fact that from the BBGKY hierarchy it is easy, essentially by
integrating over the velocities, to derive the YBG\ hierarchy which relates
the pair distribution function to the triplet distribution function, the
triplet to the quartic distribution function, etc. Some sort of closure
hypothesis, like the neglect of three body correlations, then yields a
closed equation for the pdf, see e.g.\cite{HansenMcdonald},\cite{Reichl}.
However, away from equilibrium, the distribution of velocities is not known
and this procedure cannot be followed in detail. One approach to
nonequilibrium fluid structure is that of Hess and coworkers\cite
{Rainwater83},\cite{Hess85}. The object there was to describe sheared
colloidal suspensions as well as computer simulations of sheared simple
fluids. The theory, based on a kind of relaxation approximation, is
phenomenological and will be considered in more detail below. Fluctuating
hydrodynamics offers another possibility\cite{RonisShear},\cite
{Lutsko_Fluctuations},\cite{ErnstGranularStructure} but is restricted to
length-scales much larger than atomic length scales. Another approach,
developed about a decade later by Eu et al\cite{Gan92},\cite{Gan92b}, was to
try to generalize the equilibrium integral equations, like the Kirkwood and
the Percus-Yevick equations , to nonequilibrium fluids. For the reasons just
mentioned, the only way to carry out this program was to start with an
ansatz for the n-body distribution which allows the development to parallel
that of the equilibrium equations. Whether or not this is a good approach is
difficult to assess since the approximations made are uncontrolled, but it
has been previously shown\cite{submitted2PRL}, and will be argued in more
detail below, that velocity correlations are the crucial determinant of
atomic-scale deviations from equilibrium structure and, it would follow,
that approaches which do consider them may be missing an essential
contribution to the structure. Indeed, the role played by velocity
correlations in determining one aspect of the structure of sheared simple
fluids, namely the value and angular dependence of the pair distribution
function at contact has already been demonstrated\cite{Lutsko96} and the
purpose of this paper is to show that the knowledge of such velocity
correlations can be used to extend equilibrium structural models into the
nonequilibrium domain.

In order to avoid a number of complexities associated with the spatial
anisotropy of sheared fluids, the system used here to illustrate this model
is the dissipative hard sphere model used in the study of granular fluids.
This fundamentally nonequilibrium system presents a unique opportunity for
the application and development of nonequilibrium statistical mechanics in a
system which is both of practical relevance and yet sufficiently simple to
be amenable to theoretical analysis. The basis for such analysis has
recently been formulated by Brey, Dufty and Santos\cite{GranularLiouville}
who have constructed a Liouville description of the dynamics of the
dissipative hard-sphere model. From this, Boltzmann- and Enskog-level
kinetic equations for the one-body distribution function (the dense-fluid
generalization of the Boltzmann equation) follow and their properties have
been studied theoretically, in particular, the Chapman-Enskog solution to it
has been developed, giving explicit expressions for the pressure and
transport properties of the system\cite{Garzo99}, and it is possible to
solve it numerically\cite{Santos_EnskogSimulation}. Despite the simplicity
of the dynamics, simulations have shown that this system exhibits a rich
phenomenology which is only partially understood at present (see, e.g., \cite
{Mcnamara96}). The starting point for describing the phenomenology is the
so-called homogeneous cooling state (HCS):\ since the collisions dissipate
energy, the analog of the equilibrium state is one which is spatially
homogeneous but in which the temperature decreases algebraically with time.
However, this state is relatively unstable as it is subject to a number of
forms of spontaneous symmetry breaking. The simplest example of this is the
so-called clustering instability in which sufficiently large systems,
subject to no external forces, exhibit a spatial clustering. This has been
shown to be due to a small wave-vector hydrodynamic instability\cite
{Goldhirsh93}. Small systems for which the maximum wavevector (as determined
by the size of the simulation cell) is too large, do not exhibit the
transition. Other phenomena include the ''shearing state'', in which the
system remains structurally homogeneous but long-ranged momentum
correlations develop, and ''inelastic collapse'' in which a few particles
collide with one-another many times until all relative momentum is
dissipated and the particles come to rest in contact with one another. One
motivation for trying to understand the structure of the HCS\ is that it is
needed as input to various kinetic theory calculations which could be useful
in understanding some of these features.

The remainder of this paper is organized as follows. In Section II, the
relevant elements of kinetic theory are reviewed and it is shown that the
pdf at contact can in general be explicitly calculated with the same level
of approximation as is used to get the Enskog equation. It is noted that
both the pdf at contact, as well as the pressure, can be calculated in the
HCS\ with no further approximations even though the exact solution for the
one-body distribution is not known. In Section III, the dynamics of the
dissipative hard sphere model is reformulated, through a change of
variables, so that the HCS is mapped onto a steady state. Although
mathematically, equivalent, the steady state formulation is convenient since
it avoids numerical problems associated with the rapid decay of kinetic
energy in the HCS and also allows statistical properties, like the pressure,
to be computed by replacing ensemble averages by time averages (the
assumption of ergodicity). Section IV consists of a presentation of a model
for the structure of the HCS and Section V presents a comparison of this
model to the results of simulations. The paper concludes with a discussion
of the results and their bearing on the questions raised above.

\section{Theory}

The model granular fluid consists of a collection of $N$ classical
hard-spheres of diameter $\sigma $ having positions ${\bf q}$ and momenta $%
{\bf p}$ which experience inelastic binary collisions. Between collisions,
the atoms freely stream so the state evolves according to: 
\begin{eqnarray}
m_{i}\frac{d}{dt}{\bf q}_{i} &=&{\bf p}_{i} \\
\frac{d}{dt}{\bf p}_{i} &=&{\bf F}_{i}
\end{eqnarray}
where we allow for the possibility of an external force acting on the
particles. Henceforth, we will consider only identical atoms and will use
units in which the mass is equal to one. When two particles, say $i$ and $j$%
, collide, their total momentum is unaffected but their relative momentum is
altered according to 
\begin{equation}
{\bf p}_{ij}\Rightarrow \widehat{b}_{ij}{\bf p}_{ij}\equiv {\bf p}%
_{ij}-\left( 1+\alpha \right) {\bf p}_{ij}\cdot \widehat{{\bf q}}_{ij}
\end{equation}
where, in general, relative quantities are denoted as ${\bf p}_{ij}={\bf p}%
_{i}-{\bf p}_{j}$ , unit vectors as $\widehat{{\bf q}}\equiv {\bf q}/\left| 
{\bf q}\right| $ and where this equation serves to define the momentum
transfer operator $\widehat{b}_{ij}$ . The parameter $\alpha $ is called the
coefficient of restitution and takes on values between one (elastic hard
spheres) and zero (completely dissipative inelastic collisions). Under this
dynamics, the one-body distribution function, $f_{1}(x,t)\equiv f_{1}\left( 
{\bf q},{\bf p},t\right) $ and the two-body distribution function $%
f_{2}(x_{1},x_{2},t)$ are related by the first equation of the BBGKY
hierarchy\cite{GranularLiouville} 
\begin{eqnarray}
&&\frac{\partial }{\partial t}f_{1}(x_{1},t)+{\bf p}_{1}\cdot \frac{\partial 
}{\partial {\bf q}_{1}}f_{1}(x_{1},t)  \nonumber \\
&=&\sigma ^{2}\int dx_{2}\int_{\Omega }d\widehat{{\bf \sigma }}\;\delta
\left( {\bf q}_{12}-\sigma \widehat{{\bf \sigma }}\right) \left( \widehat{%
{\bf \sigma }}\cdot {\bf p}_{12}\right) \left[ \alpha ^{-2}\widehat{b}%
_{ij}^{-1}+1\right] \Theta \left( -\widehat{{\bf q}}_{12}\cdot {\bf p}%
_{12}\right) f_{2}(x_{1},x_{2},t)  \label{2-body}
\end{eqnarray}
where the notation indicates that integration of $\widehat{{\bf \sigma }}$
is over the unit sphere, $\Theta (x)=1$ for $x>0$ and is otherwise zero and
where $\widehat{b}_{ij}^{-1}$ is the inverse of $\widehat{b}_{ij}$ and may
be easily seen to be 
\begin{equation}
\widehat{b}_{ij}^{-1}{\bf p}_{ij}\equiv {\bf p}_{ij}-\left( 1+\alpha
^{-1}\right) {\bf p}_{ij}\cdot \widehat{{\bf q}}_{ij}\widehat{{\bf q}}_{ij}
\end{equation}
Finally, the two-body distribution must satisfy the identity, a kind of
boundary condition\cite{Lutsko96}, 
\begin{eqnarray}
\delta \left( {\bf q}_{12}-\sigma \widehat{{\bf \sigma }}\right) \Theta
\left( \widehat{{\bf q}}_{12}\cdot {\bf p}_{12}\right) f_{2}(x_{1},x_{2},t)
&=&\delta \left( {\bf q}_{12}-\sigma \widehat{{\bf \sigma }}\right) \Theta
\left( \widehat{{\bf q}}_{12}\cdot {\bf p}_{12}\right) \frac{1}{\alpha ^{2}}%
\widehat{{\bf b}}_{12}^{-1}f_{2}(x_{1},x_{2},t)  \nonumber \\
&=&\delta \left( {\bf q}_{12}-\sigma \widehat{{\bf \sigma }}\right) \frac{1}{%
\alpha ^{2}}\widehat{b}_{12}^{-1}\Theta \left( -\widehat{{\bf q}}_{12}\cdot 
{\bf p}_{12}\right) f_{2}(x_{1},x_{2},t)
\end{eqnarray}
the origin of which - basically, the conservation of probability during a
collision - is discussed in the appendix. Here, I only note that this
identity is independent of, and provides information additional to, the
first BBGKY\ hierarchy equation given above. In fact, as discussed in the
appendix, it can actually be derived from the second equation of the BBGKY
hierarchy. Using an abbreviated notation, $W_{12}\equiv $ $\delta \left( 
{\bf q}_{12}-\sigma \widehat{{\bf \sigma }}\right) $, it is easily seen that
the full distribution at contact can be written as 
\begin{eqnarray}
W_{12}f_{2}(x_{1},x_{2},t) &=&W_{12}\Theta \left( -\widehat{{\bf q}}%
_{12}\cdot {\bf p}_{12}\right) f_{2}(x_{1},x_{2},t)+W_{12}\Theta \left( 
\widehat{{\bf q}}_{12}\cdot {\bf p}_{12}\right) f_{2}(x_{1},x_{2},t) 
\nonumber \\
&=&W_{12}\Theta \left( -\widehat{{\bf q}}_{12}\cdot {\bf p}_{12}\right)
f_{2}(x_{1},x_{2},t)+W_{12}\frac{1}{\alpha ^{2}}\widehat{b}_{12}^{-1}\Theta
\left( -\widehat{{\bf q}}_{12}\cdot {\bf p}_{12}\right) f_{2}(x_{1},x_{2},t)
\end{eqnarray}
The combination $\delta \left( {\bf q}_{12}-\sigma \widehat{{\bf \sigma }}%
\right) \Theta \left( -\widehat{{\bf q}}_{12}\cdot {\bf p}_{12}\right) f_{2}(%
{\bf x}_{1},{\bf x}_{2},t)$ appearing here, as well as in the collisional
term in eq.(\ref{2-body}) refers to the probability for two atoms just prior
to a collision and we refer to it as the pre-collisional part of the
distribution. Equation (\ref{bc}) thus expresses the two-body distribution
at contact solely in terms of the pre-collisional distribution. As discussed
in ref. \cite{Lutsko96}, the assumption of ''molecular chaos''\ used to
obtain the Boltzmann equation and its generalization to dense hard-sphere
fluids, the Enskog equation, is that this pre-collisional distribution can
be approximated by neglecting velocity correlations between the particles so
that one writes 
\begin{equation}
W_{12}\Theta \left( -\widehat{{\bf q}}_{12}\cdot {\bf p}_{12}\right)
f_{2}(x_{1},x_{2},t)\simeq W_{12}\Theta \left( -\widehat{{\bf q}}_{12}\cdot 
{\bf p}_{12}\right) g_{0}({\bf q}_{1},{\bf q}%
_{2})f_{1}(x_{1},t)f_{1}(x_{2},t)  \label{bc}
\end{equation}
where $g_{0}({\bf q}_{1},{\bf q}_{2};t)$ is the {\em pre-collisional }pair
distribution function (pdf)\ at time $t$ which is normally taken to be the
local-equilibrium form\cite{Beijeren79},\cite{Evans79}. Using this in eq. (%
\ref{2-body}), the generalized Enskog equation immediately results 
\begin{eqnarray}
&&\frac{\partial }{\partial t}f_{1}(x_{1},t)+{\bf p}_{1}\cdot \frac{\partial 
}{\partial {\bf q}_{1}}f_{1}(x_{1},t)  \nonumber \\
&=&\sigma ^{2}\int dx_{2}\int d\widehat{{\bf \sigma }}\;\delta \left( {\bf q}%
_{12}-\sigma \widehat{{\bf \sigma }}\right) \left( \widehat{{\bf \sigma }}%
\cdot {\bf p}_{12}\right) \left[ \alpha ^{-2}\widehat{b}_{ij}^{-1}+1\right]
\Theta \left( -\widehat{{\bf q}}_{12}\cdot {\bf p}_{12}\right) g_{0}({\bf q}%
_{1},{\bf q}_{2})f_{1}(x_{1},t)f_{1}(x_{2},t)  \label{Enskog}
\end{eqnarray}
while from the boundary condition, we get 
\begin{eqnarray}
W_{12}f_{2}(x_{1},x_{2},t) &=&W_{12}\Theta \left( -\widehat{{\bf q}}%
_{12}\cdot {\bf p}_{12}\right) g_{0}({\bf q}_{1},{\bf q}%
_{2})f_{1}(x_{1},t)f_{1}(x_{2},t)  \nonumber \\
&&+W_{12}\frac{1}{\alpha ^{2}}\widehat{b}_{12}^{-1}\Theta \left( -\widehat{%
{\bf q}}_{12}\cdot {\bf p}_{12}\right) g_{0}({\bf q}_{1},{\bf q}%
_{2})f_{1}(x_{1},t)f_{1}(x_{2},t)  \nonumber \\
&=&W_{12}g_{0}({\bf q}_{1},{\bf q}_{2})f_{1}(x_{1},t)f_{1}(x_{2},t) 
\nonumber \\
&&+W_{12}\Theta \left( \widehat{{\bf q}}_{12}\cdot {\bf p}_{12}\right)
\left( \frac{1}{\alpha ^{2}}\widehat{b}_{12}^{-1}-1\right) g_{0}({\bf q}_{1},%
{\bf q}_{2})f_{1}(x_{1},t)f_{1}(x_{2},t)  \label{final_bc}
\end{eqnarray}
thus expressing the distribution at contact as the sum of an uncorrelated
term, the first on the right, and a term expressing the corrections due to
momentum correlations. In equilibrium, the second term vanishes. Naturally,
the momenta of the atoms are correlated {\em after} a collision and, in
fact, (\ref{final_bc}) allows us to determine these correlations in a manner
consistent with the degree of approximation of the generalized Enskog
equation. Although such an evaluation might be used in a number of different
ways, I will here focus one particular application of it to the problem of
understanding the structure of the non-equilibrium state. Specifically,
integrating over momenta gives 
\begin{eqnarray}
&&W_{12}n({\bf q}_{1};t)n({\bf q}_{2};t)g({\bf q}_{1},{\bf q}_{2};t) 
\nonumber \\
&=&W_{12}n({\bf q}_{1})n({\bf q}_{2})g_{0}({\bf q}_{1},{\bf q}_{2};t) 
\nonumber \\
&&+W_{12}g_{0}({\bf q}_{1},{\bf q}_{2})\int d{\bf p}_{1}d{\bf p}_{2}\Theta
\left( \widehat{{\bf q}}_{12}\cdot {\bf p}_{12}\right) \left( \frac{1}{%
\alpha ^{2}}\widehat{b}_{12}^{-1}-1\right) f_{1}(x_{1},t)f_{1}(x_{2},t)
\label{bc2}
\end{eqnarray}
where the nonequilibrium density is $n({\bf q}_{1};t)=\int d{\bf p}%
_{1}f_{1}(x_{1},t)$. Eq. (\ref{bc2}) thus gives us an approximate evaluation
of the contribution of momentum correlations to the structure of the fluid
as expressed through the pdf. This relation was first used in ref. \cite
{Lutsko96} to characterize velocity correlations in a sheared fluid and has
recently been used in a study of velocity correlations in granular fluids
near equilibrium\cite{SotoVelCorr}.

The macroscopic balance equations for the density, momentum field $m{\bf U}$%
, and energy density $\frac{3}{2}k_{B}T$ follow from (\ref{2-body}) by
multiplying by $1$, ${\bf p}_{1}$, and $\frac{1}{2m}p_{1}^{2}$ respectively
and integrating over ${\bf p}_{1}$ to get \cite{Garzo99} 
\begin{eqnarray}
\frac{\partial }{\partial t}n+{\bf \nabla }\cdot n{\bf U} &=&0  \nonumber \\
\frac{\partial }{\partial t}U_{i}+{\bf U}\cdot {\bf \nabla }%
U_{i}+(mn)^{-1}\partial _{j}P_{ij} &=&0  \nonumber \\
\frac{\partial }{\partial t}T+{\bf U}\cdot {\bf \nabla }T+\frac{2}{3nk_{B}}%
(P_{ij}\partial _{j}U_{i}+{\bf \nabla }\cdot {\bf q}) &=&-T^{3/2}\zeta 
\label{Balance}
\end{eqnarray}
and explicit forms of the pressure tensor $P_{ij}$ and heat flux vector $%
{\bf q}$ are given in the literature\cite{Garzo99}. The source term in the
temperature equation is due to the cooling caused by the inelastic
collisions. It is easy to see that a spatially homogeneous solution to these
equations is possible with 
\begin{eqnarray}
n({\bf r},t) &=&n_{0}  \nonumber \\
{\bf U} &=&{\bf 0}  \nonumber \\
\frac{\partial }{\partial t}T &=&-T^{3/2}\zeta 
\end{eqnarray}
where $n_{0}$ is a constant. Because there is no potential energy in
hard-sphere systems, the time and energy scales are set by the temperature
and so, on dimensional grounds, it is clear that $\zeta $ is independent of
temperature: the temperature is therefore given by $T(t)=T_{0}\left[ 1+\frac{%
1}{2}\zeta T_{0}^{1/2}t\right] ^{-2}$ thus explicitly demonstrating the
cooling. A linear stability analysis of the equations (\ref{Balance})
expanded about the homogeneous cooling state shows that the state is
unstable against small wavevector fluctuations\cite{Goldhirsh93}.

The pdf at contact for HCS can also be evaluated without explicit knowledge
of the one body distribution (see the Appendix) and is 
\begin{eqnarray}
\chi \left( \widehat{{\bf \sigma }}\right)  &\equiv &\frac{1}{V}\int d{\bf q}%
_{1}d{\bf q}_{2}\;\delta \left( {\bf q}_{12}-\sigma \widehat{{\bf \sigma }}%
\right) g({\bf q}_{1},{\bf q}_{2};t)  \nonumber \\
&=&\frac{1+\alpha }{2\alpha }\chi _{0}  \label{chi}
\end{eqnarray}
where $\chi _{0}$ is defined by an analogous expression to this with $g({\bf %
q}_{1},{\bf q}_{2};t)$ replaced by $g_{0}({\bf q}_{1},{\bf q}_{2};t)$.
Finally. since it will be of use below, we quote the expression for the
collisional contribution to the pressure which is 
\begin{eqnarray}
\frac{1}{nk_{B}T}p^{c} &=&\frac{1}{3}\sum_{i=1}^{3}\frac{1}{nk_{B}T}%
P_{ij}^{c}  \nonumber \\
&=&\frac{1+\alpha }{4}\sigma ^{3}\int_{\Omega }d\widehat{{\bf \sigma }}\int
dx_{1}dx_{2}\;\delta \left( q_{12}-\sigma \right) \Theta \left( \widehat{%
{\bf q}}_{12}\cdot {\bf p}_{12}\right) \;\left( \widehat{{\bf q}}_{12}\cdot 
{\bf p}_{12}\right) ^{2}f_{1}(x_{1},t)f_{1}(x_{2},t)g_{0}({\bf q}_{1},{\bf q}%
_{2})  \nonumber \\
&=&\frac{1+\alpha }{2}\chi _{0}  \label{Pressure}
\end{eqnarray}
where the last line follows from the spatial isotropy and time invariance of
the HCS and also does not require explicit knowledge of the one-body
distribution. This is useful since an exact solution to the Enskog equation
for the one-body distribution for HCS is not known.

\section{Mapping HCS to a steady state}

Since the time scale is determined by the temperature, we can simplify the
description of this state by making a change of variables. Specifically,
define a scaled time coordinate via $ws=\ln (t/t_{0})$ where $w$ and $t_{0}$
are arbitrary constants, and the corresponding equations of motion are 
\begin{eqnarray}
\frac{d}{ds}{\bf q} &=&{\bf c}_{i}  \nonumber \\
\frac{d}{ds}{\bf c}_{i} &=&w{\bf c}_{i}  \label{scaled_eom}
\end{eqnarray}
where $\vec{c}_{i}=wt_{0}e^{ws}{\bf p}_{i}$ is the momentum (velocity) in
the new coordinates. To carry through the statistical description, we must
define a new set of probability densities. If we denote the set of positions
and velocities of atoms $x_{1}...x_{m}$ in the original system by $\Gamma
_{m}$ and of those in the scaled system by $\Gamma _{m}^{\prime }$ then the
m-body distribution will transform according to 
\begin{equation}
f_{m}(\Gamma _{m},t)=J\widetilde{f}_{m}(\Gamma _{m}^{\prime },t)
\end{equation}
where $J=\left| \frac{\partial \Gamma _{m}^{\prime }}{\partial \Gamma
_{m}^{\prime }}\right| =\left( wt_{0}e^{ws}\right) ^{mD}$ is the Jacobian of
the transformation for a D-dimensional system. We then find, e.g., that eq.(%
\ref{2-body}) becomes 
\begin{eqnarray}
&&\frac{\partial }{\partial s}\widetilde{f}_{1}(x_{1}^{\prime },s)+{\bf c}%
_{1}\cdot \frac{\partial }{\partial {\bf q}_{1}}\widetilde{f}%
_{1}(x_{1}^{\prime },s)+\frac{\partial }{\partial {\bf c}_{1}}\cdot w{\bf c}%
_{1}\widetilde{f}_{1}(x_{1}^{\prime },s)  \nonumber \\
&=&\sigma ^{2}\int dx_{1}^{\prime }dx_{2}^{\prime }\int d\widehat{{\bf %
\sigma }}\;\delta \left( {\bf q}_{12}-\sigma \widehat{{\bf \sigma }}\right)
\left( \widehat{{\bf \sigma }}\cdot {\bf c}_{12}\right) \left[ \alpha ^{-2}%
\widehat{b}_{ij}^{-1}+1\right] \Theta \left( -\widehat{{\bf q}}_{12}\cdot 
{\bf c}_{12}\right) \widetilde{f}_{2}(x_{1}^{\prime }x_{2}^{\prime },t)
\end{eqnarray}
so that the only change is that a new term appears in the streaming
operator. This term has the same form as the artificial thermostats used in
the study of shear flow\cite{Lutsko_Fluctuations}: however, here it arises
solely from a change of variables. The balance equations become 
\begin{eqnarray}
\frac{\partial }{\partial t}\widetilde{n}+{\bf \nabla }\cdot \widetilde{n}%
{\bf V} &=&0  \nonumber \\
\frac{\partial }{\partial t}V_{i}+{\bf V}\cdot {\bf \nabla }V_{i}+(m%
\widetilde{n})^{-1}\partial _{j}\widetilde{P}_{ij}-wV_{i} &=&0  \nonumber \\
\frac{\partial }{\partial t}\widetilde{T}+{\bf V}\cdot {\bf \nabla }%
\widetilde{T}+\frac{2}{3nk_{B}}(\widetilde{P}_{ij}\partial _{j}V_{i}+{\bf %
\nabla }\cdot \widetilde{{\bf q}})-2w\widetilde{T} &=&-\widetilde{T}%
^{3/2}\zeta   \label{balance2}
\end{eqnarray}
where ${\bf V}({\bf q},s)=\int d{\bf c}\;{\bf c}\widetilde{f}%
_{1}(x_{1}^{\prime },s)$, $\frac{3}{2}k_{B}\widetilde{T}({\bf q},s)=\int d%
{\bf c}\;\frac{1}{2}mc^{2}\widetilde{f}_{1}(x_{1}^{\prime },s)$, etc. Now,
the temperature of the homogeneous state is given by 
\begin{equation}
\widetilde{T}(t)=\left( \frac{2w}{\zeta }\right) ^{2}\left( 1+\left( \frac{2w%
}{\zeta \sqrt{\widetilde{T}(0)}}-1\right) e^{-wt}\right) ^{-2}
\end{equation}
so that any initial temperature will equilibrate to a final stable value of $%
\left( \frac{2w}{\zeta }\right) ^{2}$. \ Thus, the original homogeneous
cooling state is mapped by this change of variables onto a superficially
steady state. It is worth noting that if one were to return to the original
form of the dynamics and were to periodically rescale the momenta of the
atoms so as to restore the temperature to its original value, then it is
easy to see that as the time interval between rescalings goes to zero, the
resulting equations of motion can be written in the form \ref{scaled_eom}
with the constant replaced by a complicated function of the momenta.
Periodic rescaling is commonly used in simulations of sheared fluids and has
recently been employed in the simulation of HCS\cite{SotoShearingState},\cite
{SotoVelCorr}.

This mapping also makes the hydrodynamic instability apparent. If we expand (%
\ref{balance2}) about the steady state and retain terms to second order in
the gradients (i.e., the Navier-Stokes equations) and to first order in the
densities and transform to Fourier space (with wavevector ${\bf k} $), it is
easy to see that the vorticity, $\omega =\widehat{{\bf k}}\times {\bf V}$,
satisfies 
\begin{equation}
\partial _{t}\omega +\left( \nu _{0}k^{2}-w\right) \omega =0
\end{equation}
where $\nu _{0}$ is the shear viscosity evaluated at density $n_{0}$ and
temperature $\widetilde{T}_{0}=\left( \frac{2w}{\zeta }\right) ^{2}$. It is
obviously unstable for sufficiently small wavevectors and for sufficiently
large systems, we therefore expect a spontaneous shear to develop (note that 
$\nu _{0}\sim \sqrt{\widetilde{T}_{0}}$ so that the arbitrary constant, $w$,
plays no role in the stability criterion). It should also be noted that
there is a closely related instability in the total (${\bf k}={\bf 0}$)
velocity in the mapped system which, in the linear stability analysis, obeys 
\begin{equation}
\frac{\partial }{\partial t}V_{i}({\bf 0})-wV_{i}({\bf 0})=0
\end{equation}
and so is clearly unstable for all system sizes. Since we are at liberty to
choose the initial conditions to be those for which $V_{i}({\bf 0})=0$, we
will find that this is represents a minor problem in the simulations and is
of no theoretical significance.

\section{Modelling the structure of the HCS}

In this section, we will assume spatial homogeneity so that the pdf depends
only on the scalar separation between atoms. The simplest realistic model
for the structure of the equilibrium hard sphere fluid is the Percus-Yevick
approximation(see e.g., refs.\cite{HansenMcdonald},\cite{Hoye77}). This
consists of the Ornstein-Zernike equation 
\begin{equation}
h({\bf r})=c({\bf r})+\rho \int d{\bf r}^{\prime }\ c(|{\bf r}-{\bf r}%
^{\prime }|)h(r^{\prime })  \label{OZ}
\end{equation}
where $h(r)=g(r)-1$ and $c(r)$ is the direct correlation function, together
with the boundary conditions 
\begin{eqnarray}
\Theta (\sigma -r)h(r) &=&-1  \label{PY_BC} \\
\Theta (r-\sigma )c(r) &=&0  \nonumber
\end{eqnarray}
where the first condition is exact while the second defines the
approximation. Comparison with computer simulation shows that the PY
approximation for the pair distribution function is quite accurate for
separations greater that about two hard-sphere diameters but less accurate
near contact. The description of the small-separation structure can be
significantly improved by considering the Yukawa closure for the
Orstein-Zernicke equation which replaces the boundary condition on the
direct correlation function by 
\begin{equation}
\Theta (r-1)c(r)=\sum_{i=1}^{m}K_{i}\frac{e^{-v_{i}r}}{r}  \label{YC_BC}
\end{equation}
and choosing the constants $K_{i}$ and $v_{i}$ to reproduce known
properties:\ for example, taking $m=1$ and fitting the Carnahan-Starling
equation of state as calculated by both the pressure equation and the
compressibility equation. The original Mean Spherical Approximation, for
arbitrary pair potentials $\Phi (r)$, consists of requiring that $\Theta
(r-\sigma )c(r)=\Phi (r)$ where the effective hard sphere diameter is fit
according to some criterion. The PY approximation is then seen to be the MSA
for hard spheres. Equation (\ref{YC_BC})\ may therefore be viewed as either
the MSA for a potential which is the sum of Yukawas or as a general
expansion with coefficients to be fitted in which case it is termed the
Generalized MSA\ or GMSA and can be viewed as being systematic since any
function could be fitted as a sum of Yukawas.

Note that eq.(\ref{OZ})\ defines the direct correlation function and that
the first of the boundary conditions in eq.(\ref{PY_BC})\ is an exact
requirement. The only way in which this model uses the assumption that the
system being modeled is in equilibrium is through the arguments that lead to
the conclusion that the direct correlation function is short-ranged and
hence the justification for the boundary conditions in eq.(\ref{PY_BC}) and
eq.(\ref{YC_BC}). This connection to the equilibrium state is made even
weaker in a reformulation of the model due to Yuste and Santos\cite{Yuste91}%
, \cite{Yuste93b},\cite{Yuste94}. They begin by noting that the Laplace
transform of the quantity $rg(r)$, in the PY approximation, is naturally
written as 
\begin{eqnarray}
G(t) &\equiv &\int_{0}^{\infty }dr\ e^{-sr}rg(r)  \nonumber \\
&=&\frac{tF(t)e^{-t}}{1-12\eta F(t)e^{-t}}  \label{GT}
\end{eqnarray}
with 
\begin{equation}
F(t)=\frac{1+A_{1}t}{S_{0}+S_{1}t+S_{2}t^{2}+S_{3}t^{3}}  \label{FT}
\end{equation}
They go on to point out that given the second equality of eq.(\ref{GT}) and
making a Pad\'{e}' approximation for the function $F(t)$ one can deduce the
correct order of the numerator and denominator of $F(t)$ as well as the PY
expressions for the coefficients based solely on the asymptotic properties
of the pdf. Specifically, they note that a)$g(r)$ at contact is given by $%
g(\sigma )=$ $\lim_{t\rightarrow \infty }t^{2}F(t)$ thus fixing the relative
number of terms in the numerator and denominator; b)\ $\lim_{r\rightarrow
\infty }g(r)=1$ implies that $G(t)%
\mathrel{\mathop{\rightarrow }\limits_{t\rightarrow 0}}%
t^{-2}$ and c)\ the fact that the static structure factor, given by $%
S(q)=\lim_{t\rightarrow iq}%
\mathop{\rm Re}%
\left( tG(t)\right) $, is finite at $q=0$ implies, given the previous
condition, that for small $t$, $G(t)=t^{-2}+o(1)$. (The last condition is
equivalent to assuming that long-range correlations do not exist.)\ The
minimal approximant satisfying these conditions is that given in eq.(\ref{FT}%
) with the PY value for the coefficients. They also note that the extension
of the Pad\'{e}' approximant to include one more term in both the numerator
and denominator is exactly equivalent to the one-Yukawa closure while the
further extension of the approximation corresponds to a closure consisting
of a sum of Yukawas. This method is also shown to give the PY solution for
sticky hard spheres as well as the exact structure for both ordinary and
sticky hard spheres in one-dimension. A straight forward extension of these
ideas has also been used to model the square-well fluid. Thus, in this
formulation, the PY form of $G(t)$ is taken as an ansatz characteristic of
hard-core systems and the function $F(t)$ modeled as a Pad\'{e}' approximant
subject to whatever knowledge exists about the structure.

With this justification in mind, we consider the application of this
approach to the nonequilibrium HCS. In equilibrium, the next inclusion of an
additional term in the numerator and denominator of $F(z)$ introduces two
new parameters which are used to fit a known equation of state (normally the
Carnahan-Starling equation of state) through both the pressure equation and
the compressibility equation. The pressure equation for hard spheres in
three dimensions reads\cite{HansenMcdonald} 
\begin{equation}
\frac{p}{nk_{B}T}=1+4\eta \chi _{eq}
\end{equation}
and so allows to calculate the pdf at contact, recall $\chi _{eq}=g(\sigma )$%
, from the equation of state. In a nonequilibrium state, the collisional
boundary condition can be used, together with assumption of molecular chaos,
to give the same information. The equilibrium compressibility equation is 
\begin{eqnarray}
\left( \frac{\partial }{\partial \rho }\beta P\right) ^{-1} &=&1+\rho \int d%
\vec{r}\ (g(r)-1)  \nonumber \\
&=&1+\lim_{t\rightarrow 0}\left( G(t)-\frac{1}{t^{2}}\right)
\end{eqnarray}
and for this, there is no obvious substitute for the nonequilibrium state.
With nothing to use in its place, I will continue to apply this even in the
nonequilibrium state, calculating the pressure from eq.(\ref{Pressure}),
which might be viewed as a local-equilibrium approximation. In fact, the
resulting model is relatively insensitive to the value used for the pressure
since this only fixes the area of the structure function whereas the results
are quite sensitive to the value of the pdf at contact. Using $g_{0}({\bf q}%
_{1},{\bf q}_{2};t)\simeq g_{eq}(q_{12})$ as is normally done in Enskog
theory, the model is given by 
\begin{equation}
F(t)=-\frac{1}{12\eta }\frac{1+A_{1}t+A_{2}t^{2}}{%
S_{0}+S_{1}t+S_{2}t^{2}+S_{3}t^{3}+S_{4}t^{4}}
\end{equation}
with 
\begin{eqnarray}
S_{0} &=&1  \nonumber \\
S_{1} &=&A_{1}-1  \nonumber \\
S_{2} &=&A_{2}-A_{1}+\frac{1}{2}  \nonumber \\
S_{3} &=&-A_{2}+\frac{1}{2}A_{1}-\frac{1+2\eta }{12\eta }  \nonumber \\
S_{4} &=&\frac{1}{2}A_{2}-\left( \frac{1+2\eta }{12\eta }\right) A_{1}+\frac{%
2+\eta }{24\eta }  \nonumber \\
A_{1} &=&\frac{1}{2}+\sqrt{\frac{1}{12\eta }\frac{\left( \eta -1\right)
^{2}-Z\left( 6\eta g(1;\alpha )+1\right) }{\left( 2+\eta \right)
-2g(1;\alpha )\left( \eta -1\right) ^{2}}}  \nonumber \\
A_{2} &=&g(1;\alpha )\frac{\left( 1+2\eta \right) A_{1}-\frac{1}{2}\left(
2+\eta \right) }{1+6\eta g(1;\alpha )}
\end{eqnarray}
and 
\begin{equation}
Z\equiv \left( \frac{\partial }{\partial \rho }\beta P\right) ^{-1}=\left( 1+%
\frac{1+\alpha }{2}\left( Z_{eq}-1\right) \right) ^{-1}
\end{equation}
with the Carnahan-Starling expression\cite{HansenMcdonald} 
\begin{equation}
\chi _{eq}=\frac{1-\frac{1}{2}\eta }{\left( 1-\eta \right) ^{3}}
\end{equation}
together with eqs.(\ref{Pressure}) and (\ref{chi})\ completes the model. \
This then reduces to the GMSA in equilibrium and can be seen as its natural
generalization to a spatially isotropic nonequilibrium state.

\section{Molecular Dynamics Simulations}

To determine the structure of the model granular system, I have performed
molecular dynamics simulations of a small systems of 108 and 500 particles
in 3 dimensions governed by the steady-state dynamics described by eq.(\ref
{scaled_eom}) and subject to periodic boundary conditions. The density was
taken to be $n^{\ast }=0.5$: high enough that finite density effects are
important but low enough that the Enskog approximation is expected to be
valid. The choice to simulate the steady-state dynamics, rather than to
simulate the ''real'' dynamics of the cooling system, was made on the basis
that the systems cool very rapidly so that the time scales involved in the
simulations become very large, the velocities and energies very small and
numerical inaccuracies due to round-off error are a significant problem.
This could be dealt with by periodically rescaling the velocities (i.e.,
redefining the time unit as in \cite{SotoShearingState},\cite{SotoVelCorr})
but it is far more elegant and efficient to directly simulate the
steady-state dynamics. Furthermore, the changes needed to implement this
starting with a code for simulating equilibrium hard spheres are minimal.
One point that does require attention is the instability with respect to the
total momentum. Even if initial conditions are chosen so that the total
momentum is zero at the start of the simulation, round-off errors lead to
the spontaneous appearance of a non-zero total momentum which then quickly
goes due to the instability. This effect is, however, benign and is easily
suppressed by calculating the total momentum during each propagation step
and subtracting $(1/N)$ of its value from the momentum of each particle.

The starting point for the simulations was an equilibrium configuration of
velocities and positions. The value of the thermostat constant, $w$, was set
arbitrarily. For each value of $\alpha $, the simulations were
''equilibrated'' for a total of $3\cdot 10^{6}$ collisions and then
statistics gathered over a second series of $3\cdot 10^{6}$ collisions . To
obtain steady-state averages of one-body properties, quantities were
time-averaged over periods of $10^{4}$ collisions throughout the
simulations; these samples were then treated as statistically independent
estimates and their average and standard error computed\cite{Erpenbeck83}.
The errors in all quantities reported below are found to be small, less than
1\%. The determination of collisional effects, such as in the pressure and
the pdf at contact, is somewhat different. For any collisional quantity of
the form 
\begin{equation}
A\equiv \sum_{i<j}A(x_{i},x_{j})\delta (q_{ij}-1)
\end{equation}
the ergodic assumption gives 
\begin{eqnarray}
\left\langle A\right\rangle  &=&\frac{1}{T}\int_{0}^{T}dt\ A(t)  \nonumber \\
&=&\sum_{i<j}\frac{1}{T}\int_{0}^{T}dt\ A(x_{i},x_{j})\delta (q_{ij}-1) 
\nonumber \\
&=&\sum_{i<j}\frac{1}{2T}\int_{0}^{T}dt\ \delta (\tau _{ij}-1)\left( \frac{%
A(ij)}{\left| {\bf q}_{ij}\cdot {\bf p}_{ij}\right| }+\frac{A(x_{i}^{\prime
},x_{j}^{\prime })}{\left| {\bf q}_{ij}\cdot {\bf p}_{ij}^{\prime }\right| }%
\right) 
\end{eqnarray}
where the first two lines integrate the total time dependence of the
function $A$. The third line\ follows from a change of variable in the
delta-function $\vec{p}_{ij}$($\vec{p}_{ij}^{\prime }$) is the relative
momentum of the colliding pair immediately before(after) the collision and $%
\tau _{ij}$ is the time at which the pair (i,j) collide (which could be
imaginary or outside the range of integration indicating in either case that
they do not collide). The last expression obviously reduces to a sum over
collisions: 
\begin{equation}
\left\langle A\right\rangle =\frac{1}{2T}\sum_{collisions,\gamma }\left( 
\frac{A(\gamma )}{\left| {\bf q}_{\gamma }\cdot {\bf p}_{\gamma }\right| }+%
\frac{A(\gamma ^{\prime })}{\left| {\bf q}_{\gamma }\cdot {\bf p}_{\gamma
}^{\prime }\right| }\right) _{t_{\gamma }}
\end{equation}
where $\gamma $ represents the colliding pair and which is the form used to
evaluate the collisional part of the pdf at contact, $\frac{N(N-1)}{2V}4\pi
g(1)=\left\langle \sum_{i<j}\delta (q_{ij}-1)\right\rangle $. Using $\left| 
{\bf q}_{\gamma }\cdot {\bf p}_{\gamma }^{\prime }\right| =\alpha \left| 
{\bf q}_{\gamma }\cdot {\bf p}_{\gamma }\right| $, this becomes 
\begin{equation}
g(1)=\frac{1}{2\pi n}\left( \frac{\alpha +1}{2\alpha }\right) \frac{1}{T}%
\sum_{collisions,\gamma }\frac{1}{\left| {\bf q}_{\gamma }\cdot \vec{p}%
_{\gamma }\right| }\Theta \left( -{\bf q}_{\gamma }\cdot {\bf p}_{\gamma
}\right) 
\end{equation}
while for the pressure one finds 
\begin{equation}
p^{c}=\frac{\left( 1+\alpha \right) }{2T}\sum_{collisions,\gamma }\left| 
{\bf q}_{\gamma }\cdot {\bf p}_{\gamma }\right| \Theta \left( -{\bf q}%
_{\gamma }\cdot {\bf p}_{\gamma }\right)   \label{md-pressure}
\end{equation}
and in both cases, the step function indicates that the expression is
evaluated with the pre-collisional momenta. Finally, we present below
determinations of the pdf for finite separations. These are determined in
the obvious way by looping over all pairs of atoms and creating a histogram
of the separations. The bin size used was 0.025 (hard sphere diameters) and
these were compiled every $10^{4}$ collisions and the results averaged to
obtain the final histogram.

\begin{figure}[t]
\begin{center}
\leavevmode
\epsfxsize=5in 
\epsfysize=5in
\epsfbox{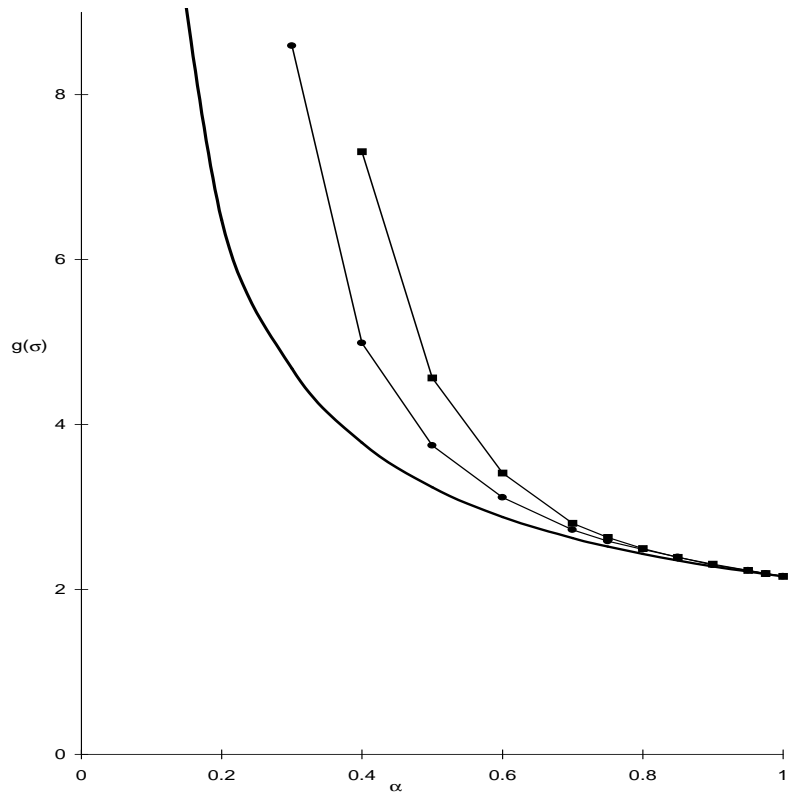}
\end{center}
\caption{The pdf at contact for n*=0.5 from simulation of 108 atoms (circles) \
500 atoms (squares) and from eq.(\ref{chi}). The lines between the simulation data
are only a guide to the eye.}
\label{fig1}
\end{figure}

Figure \ref{fig1} shows the pdf at contact as determined from the
simulations and from the collisional boundary condition. For $\alpha >0.6$,
the agreement is seen to be good but it becomes worse values below this:
furthermore, there appears to be a strong number dependence to the results
with the larger system diverging more rapidly from the prediction. In both
cases, simulations are only possible for $\alpha $ above some threshold:
below this, the simulation code fails due to the time between collisions
becoming smaller and smaller until the machine precision is reached. A
detailed analysis of the sequence of collisions shows that this is due to a
small number of particles with virtually no momentum relative to one another
colliding over and over again - in other words, this is the phenomena of
elastic collapse described by McNamara and Young\cite{Mcnamara94}. The
threshold for the collapse is in the range $0.3<\alpha <0.4$ for the 500
particle system and $0.2<\alpha <0.3$ for the 108 particle system. In both
cases, prior to the collapse, the value of the pdf at contact is several
times the highest values shown in figure \ref{fig1}.

\begin{figure}[t]
\begin{center}
\leavevmode
\epsfxsize=5in 
\epsfysize=5in
\epsfbox{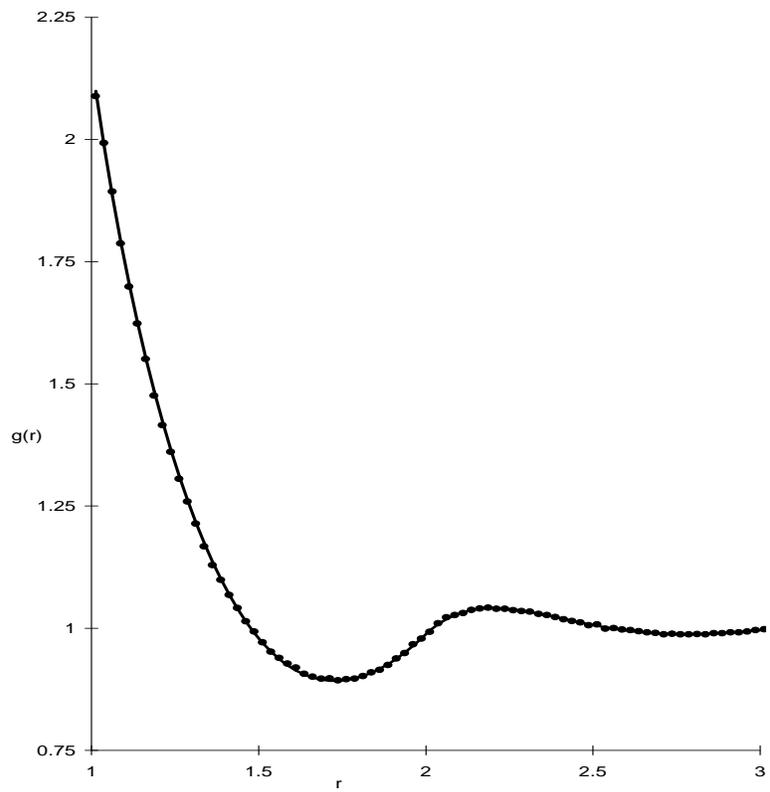}
\end{center}
\caption{The pdf at equilibrium ($\alpha =1$) for n*=0.5 as determined from
simulation of 500 atoms (circles) and from the(GMSA)\ model (curve).}
\label{fig2}
\end{figure}

\begin{figure}[t]
\begin{center}
\leavevmode
\epsfxsize=5in 
\epsfysize=5in
\epsfbox{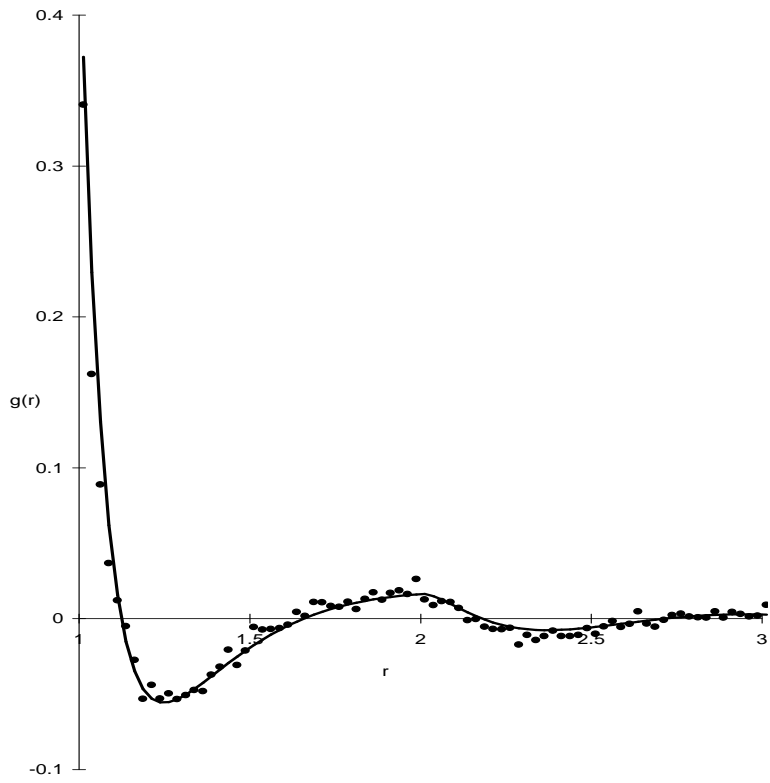}
\end{center}
\caption{The nonequilibrium part of the pdf, $\delta g(r)=g(r;\alpha)-g(r;1)$,
 for n*=0.5 atoms with $\alpha =0.7$ as
determined from simulation of 108 atoms (circles) and from the model (curve).}
\label{fig3}
\end{figure}

\begin{figure}[t]
\begin{center}
\leavevmode
\epsfxsize=5in 
\epsfysize=5in
\epsfbox{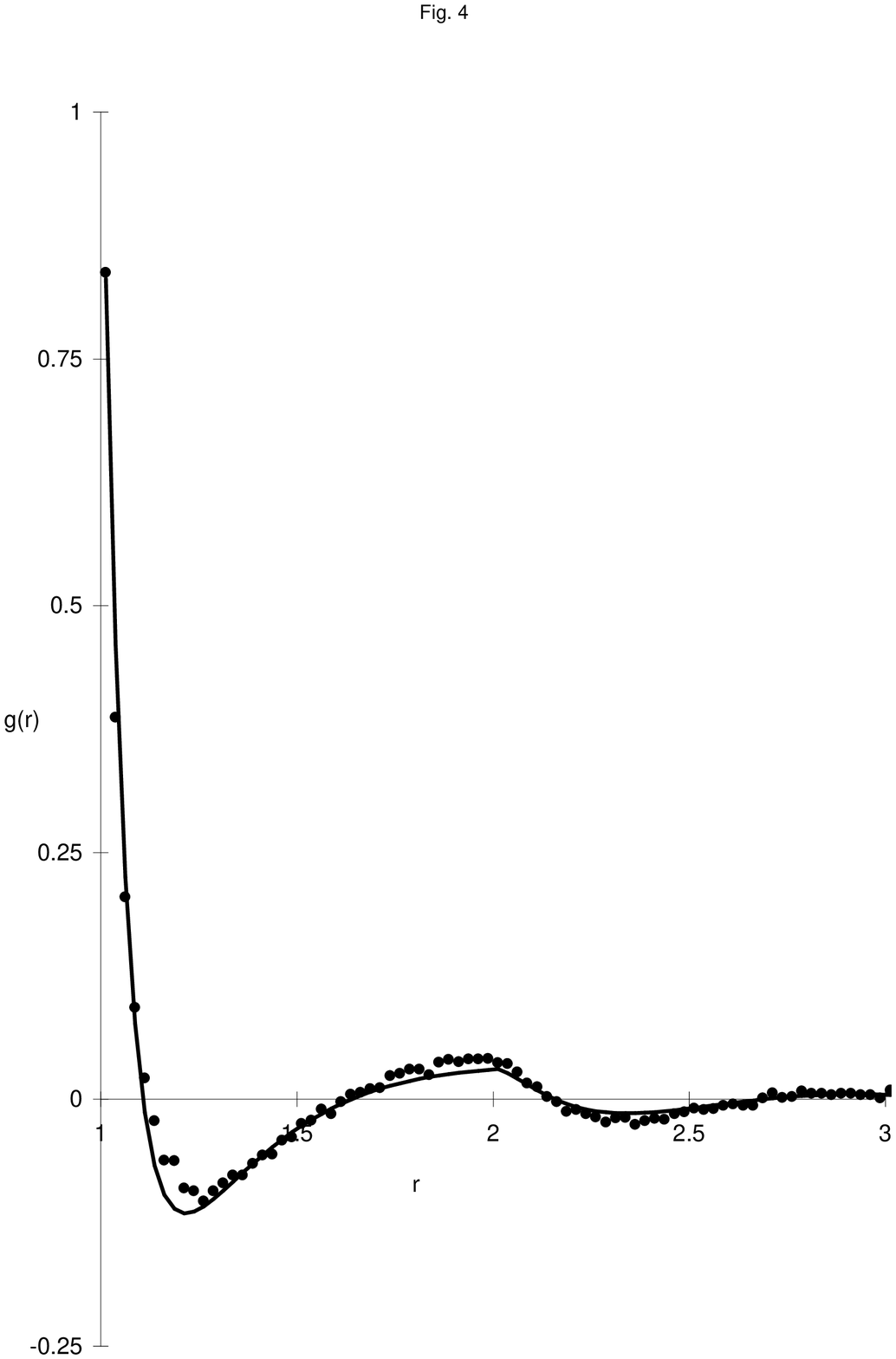}
\end{center}
\caption{Same as fig. 3 for $\alpha =0.5$ and  108 atoms.}
\label{fig4}
\end{figure}

\begin{figure}[t]
\begin{center}
\leavevmode
\epsfxsize=5in 
\epsfysize=5in
\epsfbox{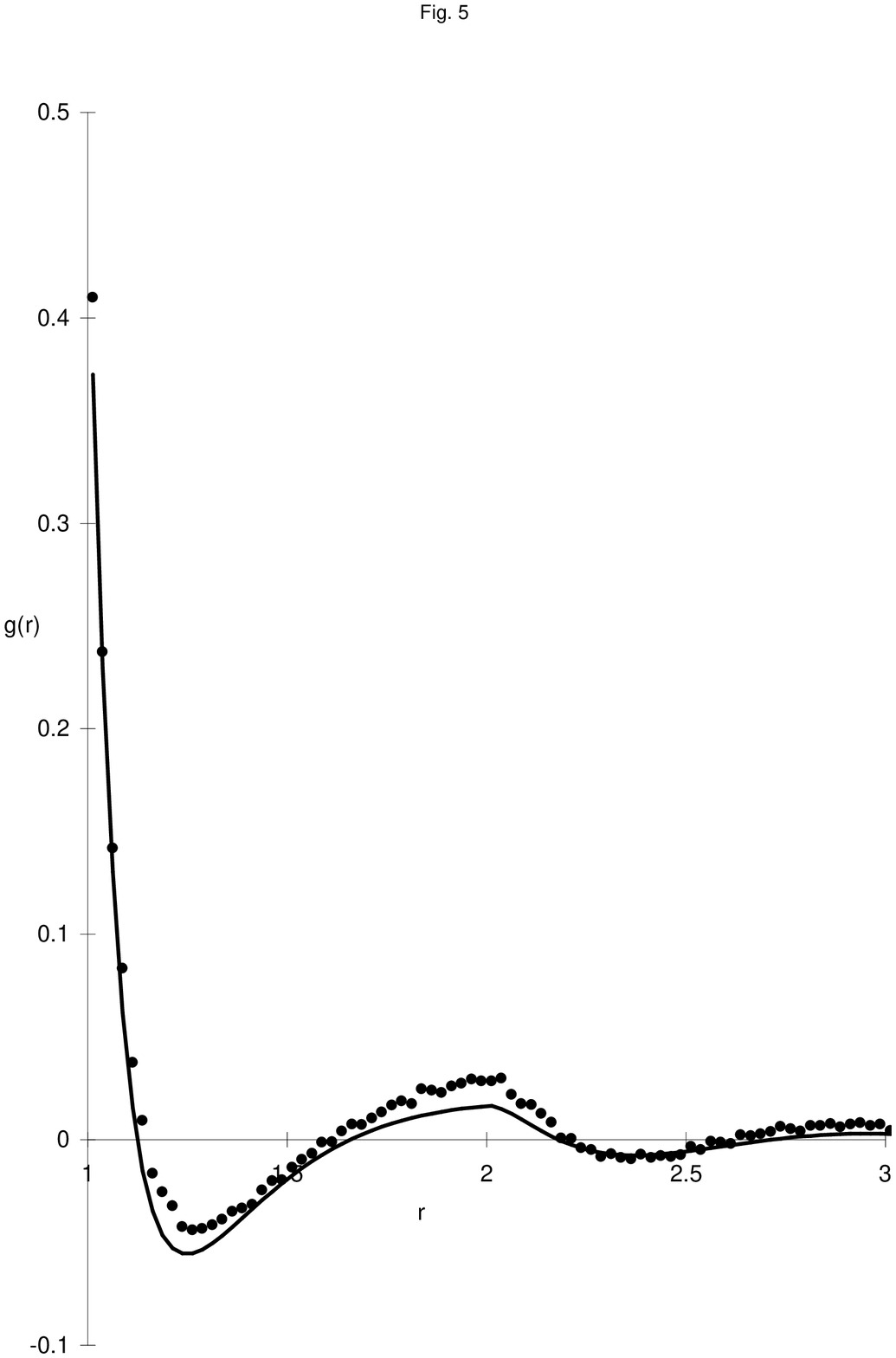}
\end{center}
\caption{Same as fig. 3 for $\alpha =0.7$ and 500 atoms.}
\label{fig5}
\end{figure}

\begin{figure}[t]
\begin{center}
\leavevmode
\epsfxsize=5in 
\epsfysize=5in
\epsfbox{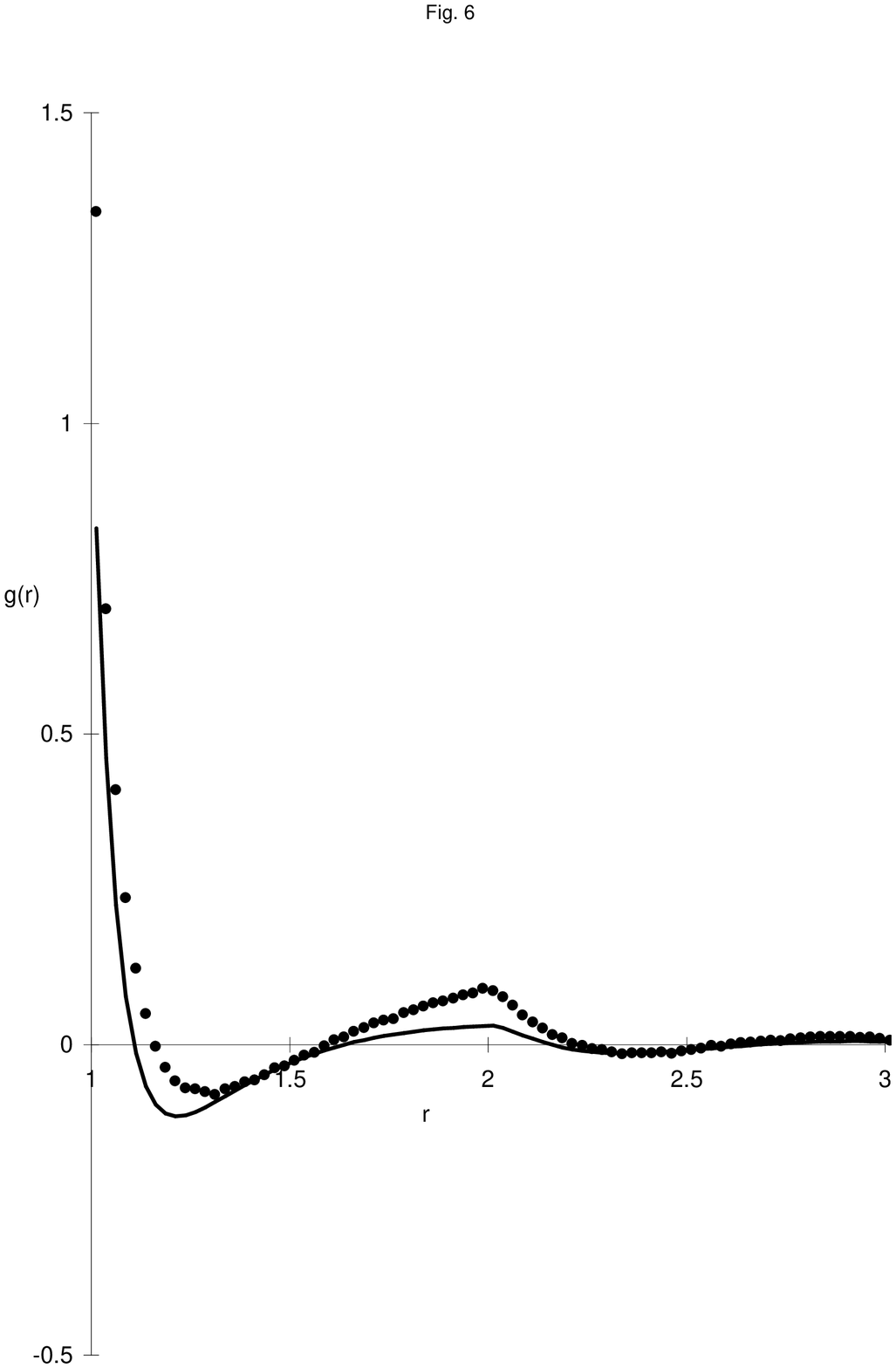}
\end{center}
\caption{Same as fig. 3 for $\alpha =0.5$ and 500 atoms.}
\label{fig6}
\end{figure}

Similar behavior is seen in the pair distribution function. For reference,
Fig. \ref{fig2} shows the pdf at equilibrium as determined from the 500 atom
simulations and from the model: the agreement is seen to be excellent as is
also the case for the data coming from the 108 atom system. Figures \ref
{fig3} and \ref{fig4} show the nonequilibrium part of the pdf (i.e., $%
g(r)-g_{eq}(r)$) for the 108 atom system as determined by simulation and by
from the nonequilibrium GMSA for $\alpha =0.7$ and $0.5$ respectively and in
both cases, the extended MSA is seen to give a good quantitative description
of all features of the nonequilibrium structure. (In fact, since the MD
results are, by their nature, binned, the theoretical curve is obtained by
integrating the model pdf over bins of the same size and position as used in
the simulations.) Figures \ref{fig5} and \ref{fig6} show that, not
surprisingly, the agreement is less good for the 500 atom system,
particularly at the smaller value of $\alpha $.

\begin{figure}[t]
\begin{center}
\leavevmode
\epsfxsize=5in 
\epsfysize=5in
\epsfbox{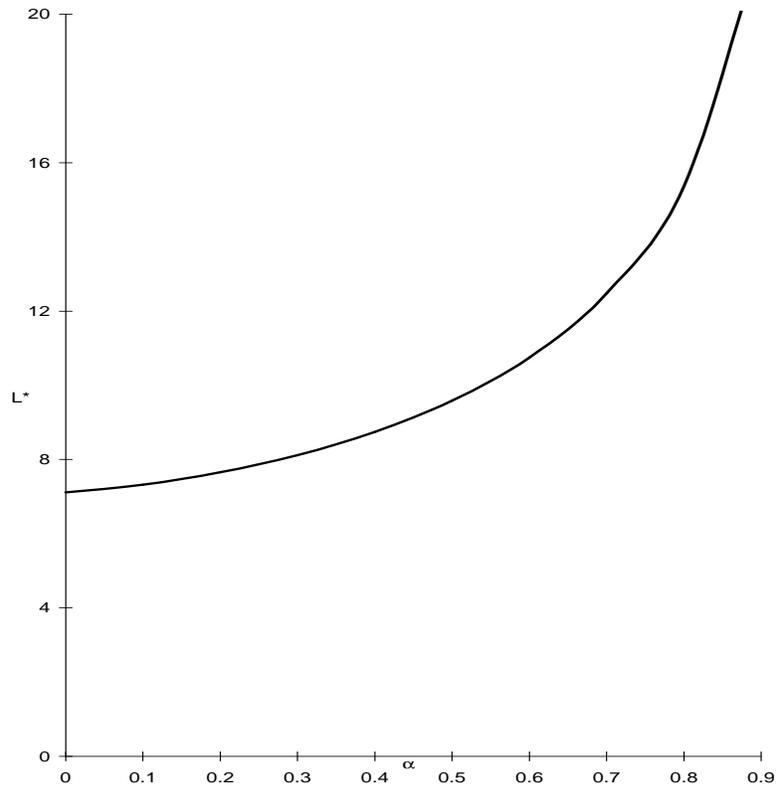}
\end{center}
\caption{The critical system sizes as a function of $\alpha $ for n*=0.5. Systems
falling below the curve are unstable.}
\label{fig7}
\end{figure}

\begin{figure}[t]
\begin{center}
\leavevmode
\epsfxsize=5in 
\epsfysize=5in
\epsfbox{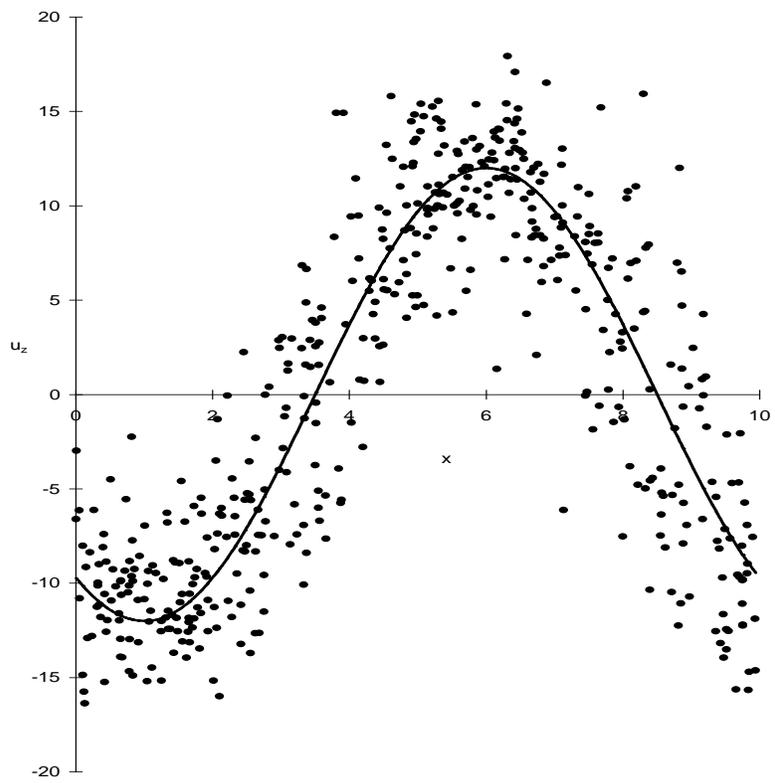}
\end{center}
\caption{A snapshot of the 500 atom system:\ the horizontal axis is the position
along the x-axis of the simulation, the vertial axis shows the momentum
along the z-direction. The curve is a sin-function fitted to the data.}
\label{fig8}
\end{figure}

The disparity between the results for the two systems is much greater than
one finds in equilibrium and suggests that a qualitative difference between
them. One obvious possible source of such a difference is the hydrodynamic
instability discussed above. Using the values for the transport coefficients
given by \cite{Garzo99}, one finds the critical size curve shown in Fig. \ref
{fig7} which indicates that the 108 atom system is always stable but that
the 500 atom system becomes unstable around $\alpha <0.6$ however, knowledge
of the transport coefficients at small $\alpha $ is only approximate so
these numbers may only be indicative of the position of the instability.
Nevertheless, the importance of the instability in the larger system is
easily confirmed and Fig. \ref{fig8}, showing the velocities in one
direction versus the positions along another as taken from a snapshot of the
500 atom system with $\alpha =0.5$, shows a spontaneously formed shearing
profile. In a larger system, this would manifest itself in the form of
vortices. Further evidence of the instability can be found in the kinetic
contributions to the pressure tensor where, beginning at $\alpha =0.7$ in
the 500 atom system, an oscillation develops whereby a large fraction, on
the order of $2/3$, of the kinetic energy is concentrated in first one
component of the pressure tensor and then another indicating that
macroscopic flows are forming. The obvious interpretation is that around $%
\alpha =0.7$ the shear mode is soft or unstable. There is no evidence of
such an unstable mode for any value of at $\alpha $ in the 108 atom system.
This picture is thus in qualitative agreement with the predictions based on
the calculations described above and has recently been observed in other
studies\cite{SotoShearingState}.

\begin{figure}[t]
\begin{center}
\leavevmode
\epsfxsize=5in 
\epsfysize=5in
\epsfbox{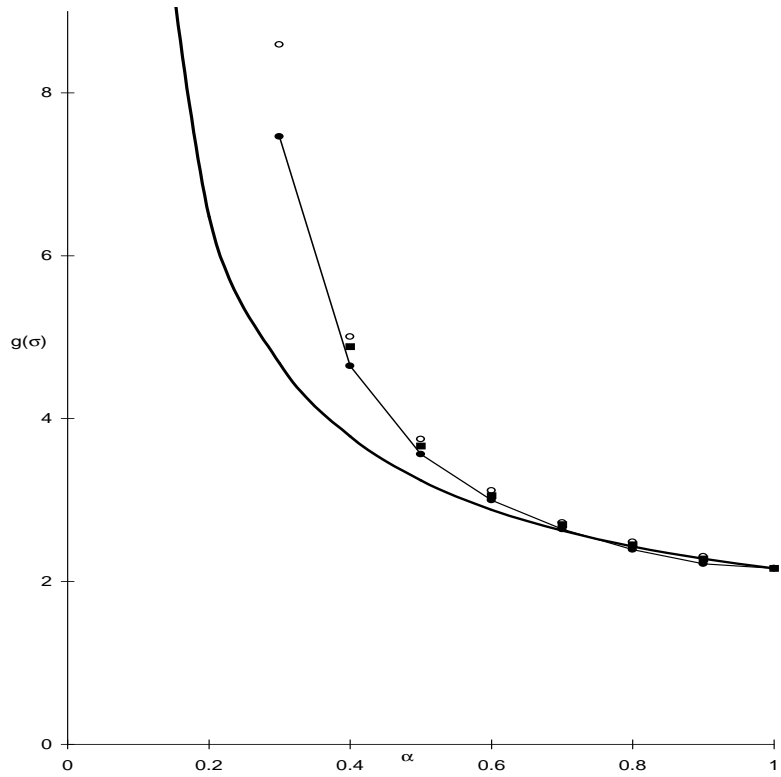}
\end{center}
\caption{The pdf at contact for n*=0.5 from the original unconstrained simulation
of 108 atoms (open circles), the constrained simulation of 108 atoms (circles)
and \ the constrained simulation of 500 atoms (squares) and from eq.(\ref
{chi}) (line). The lines between the simulation data
are only a guide to the eye.}
\label{fig9}
\end{figure}

\begin{figure}[t]
\begin{center}
\leavevmode
\epsfxsize=5in 
\epsfysize=5in
\epsfbox{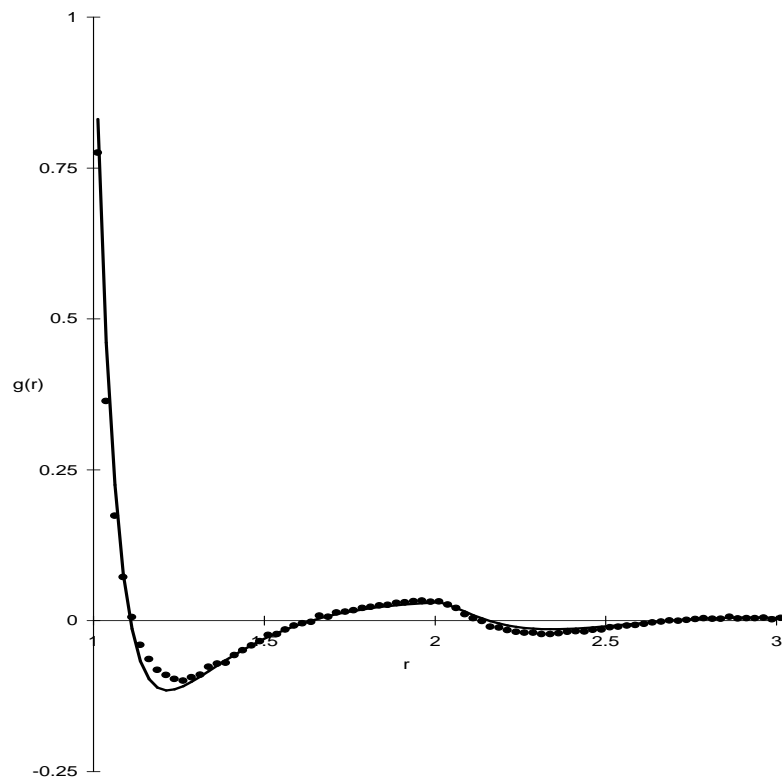}
\end{center}
\caption{Same as fig. 6 for $\alpha =0.7$ and 500 atoms and showing data from the constrained simulation.}
\label{fig10}
\end{figure}

We can suppress the instability in a crude way by periodically adjusting the
velocities of the atoms. In these ''constrained'' simulations we interpose
correction whereby after every 100 collisions, we calculate the amplitude of
the longest wavelength Fourier modes of the systems (i.e., ${\bf A}_{l}=%
\frac{2}{N}\sum_{i=1}^{N}{\bf c}_{i}\cos {\bf k}_{l}\cdot {\bf q}_{i}$ for $%
{\bf k}_{1}=\left( 2\pi /L\right) \widehat{{\bf x}}$, etc.) \ and we then
subtract the mode from each atom's velocity ( ${\bf c}_{i}\rightarrow {\bf c}%
_{i}-\sum_{l=1}^{3}{\bf A}_{l}\cos {\bf k}_{l}\cdot {\bf q}_{i}$). This is a
crude procedure in that the amplitudes of the modes are only approximately
set to zero and it also has the effect of removing kinetic energy from the
system (which is, however, masked by the input of kinetic energy coming from
the equations of motion). A more elegant procedure could be devised based on
standard nonequilibrium molecular dynamics techniques such as Gauss'
principle of least constraint\cite{GaussPrinciple}, but as the present
purpose is only to control the unstable mode, the crude method was deemed
sufficient. As shown in Fig. \ref{fig9}, the result is to give better
agreement in the measured value of the pdf at contact between the two
systems while having relatively little effect in the smaller system except
at the highest values of alpha. Figure \ref{fig10} shows that the pdf as
determined from the constrained simulation of the larger system is in
considerably better agreement with the model.

\section{Discussion}

The main purpose of this paper has been to show that the GMSA\ can be
extended to nonequilibrium systems by replacing the equilibrium input
required by the GMSA with accessible nonequilibrium information coming from
the collisional boundary condition and, incidently, to elucidate the
atomic-scale structure of the HCS of granular fluids. In order to compare
the predicted values of the pdf at contact and the pressure with the results
of simulations, the equations of motion of the dissipative hard sphere
system were mapped onto those that describe a steady state thus allowing us
to use standard methods of steady state simulation such as the replacement
of ensemble averages by time averages. The comparisons with molecular
dynamics simulations also show that the pair distribution function at
contact can be used as a signal of the onset of elastic collapse - its value
steadily diverges from the predicted value as the elastic collapse threshold
is approached and its value in the simulations that feature the collapse is
very large.

\begin{figure}[t]
\begin{center}
\leavevmode
\epsfxsize=5in 
\epsfysize=5in
\epsfbox{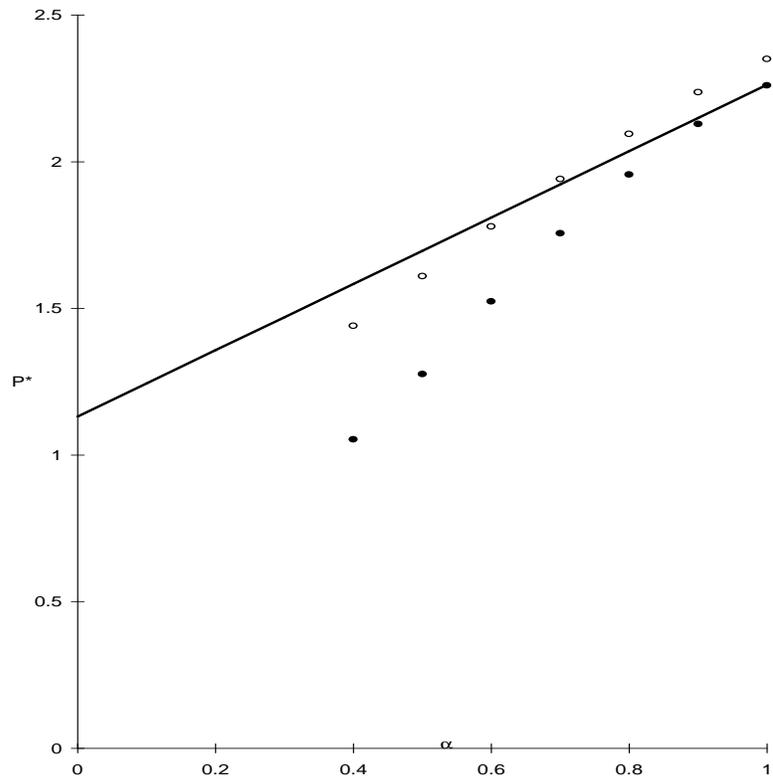}
\end{center}
\caption{The collisional contribution to the reduced pressure,$P/nk_{B}T$, for
n*=0.5 as determined from the unconstrained 108 atom simulation (diamonds)
and from eq.(\ref{Pressure}).}
\label{fig11}
\end{figure}

The significant differences observed between the 108 and 500 atom systems
were seen to be largely due to the soft hydrodynamic modes present in the
larger system. Nevertheless, even when these are accounted for, there
remains a significant deviation of the simulation results from the various
predictions of the Enskog theory. It is tempting to conclude that this is
due to a poor estimation of the quantity $\chi _{0}$, for which we use the
equilibrium value, but examination of the pressure shows that the answer
cannot be this simple:\ the pdf at contact would require a {\em larger}
value for $\chi _{0}$ that increases with $\alpha $ in order to be in
agreement with the simulations whereas Fig. \ref{fig11} shows that pressure
would require a {\em smaller} value that decreased with $\alpha $ (in both
the unconstrained and constrained simulations). We thus conclude that the
deviations are due to the Enskog approximation itself and could probably be
described via mode coupling. Nevertheless, that $\chi _{0}$ should depend on 
$\alpha $ is intuitively clear:\ atoms moving slowly away from a collision
will be more likely to be knocked, by a third atom, into a second collision
with one another leading to such a dependence.

As mentioned in the Introduction, one phenomenological approach to the
description of nonequilibrium structure is that of Hess and coworkers\cite
{Rainwater83},\cite{Hess85}. In its simplest form, this reduces to a
relaxation model for the nonequilibrium contribution to the pdf: 
\begin{equation}
\frac{\partial }{\partial t}g({\bf r},t)+{\bf v}({\bf r},t)\cdot {\bf \nabla 
}g({\bf r},t)=\tau ^{-1}\left( g({\bf r},t)-g_{0}({\bf r})\right)
\end{equation}
where $\tau $ is a relaxation time and $g_{0}({\bf r})$ is taken to be the
equilibrium pdf. It is clear that for a homogeneous steady state with no
flow this gives the trivial result that $g({\bf r},t)=g_{0}({\bf r})$.
However, this model is intended only as a simplification of a more complex
model given by 
\begin{equation}
\frac{\partial }{\partial t}g({\bf r},t)+{\bf v}({\bf r},t)\cdot {\bf \nabla 
}g({\bf r},t)+D{\bf \nabla }\cdot \left( g_{0}({\bf r},t){\bf \nabla }\left(
g({\bf r},t)/g_{0}({\bf r})\right) \right) =0
\end{equation}
where $D$ is a kind of diffusion constant. Again, for a homogeneous steady
state with no flow, only the last term survives so that this reduces to 
\begin{equation}
g_{0}(r)\frac{1}{r^{2}}\frac{d}{dr}r^{2}\frac{d}{dr}\frac{g(r)}{g_{0}(r)}+%
\frac{dg_{0}(r)}{dr}\frac{d}{dr}\frac{g(r)}{g_{0}(r)}=0
\end{equation}
which has the solution 
\begin{eqnarray}
g(r) &=&g_{0}(r)+\left( \frac{g(1)}{g_{0}(1)}-1\right) \frac{f(r)}{f(1)}%
g_{0}(r)  \nonumber \\
f(r) &=&\int_{r}^{\infty }dr\frac{r^{-2}}{g_{0}(r)}
\end{eqnarray}
where we have used the boundary condition $\lim_{r\rightarrow \infty }g(r)=1$%
. Although this has an interesting structure, the function $f(r)$ is
positive-definite so that for HCS the difference $g(r)-g_{0}(r)$ is also
which precludes a description of the oscillatory nature of the differences
found in the simulation.

Another approach is the theory of Eu et al\cite{Gan92},\cite{Gan92b}, also
mentioned in the introduction. This theory is based on an ansatz for the
N-body nonequilibrium distribution and takes the form of an integral
equation: 
\begin{eqnarray}
\ln y({\bf q}_{1},{\bf q}_{2};t) &=&n\int d{\bf q}_{3}\;f_{NE}\left( {\bf q}%
_{1},{\bf q}_{3};t\right) y({\bf q}_{1},{\bf q}_{3};t)  \nonumber \\
&&\times \left[ y({\bf q}_{2},{\bf q}_{3};t)\left( 1+f_{NE}\left( {\bf q}%
_{2},{\bf q}_{3};t\right) \right) -1\right] 
\end{eqnarray}
where 
\begin{eqnarray}
g\left( {\bf q}_{1},{\bf q}_{2};t\right)  &=&\exp \left[ -V_{NE}\left( {\bf q%
}_{1},{\bf q}_{2};t\right) \right] y({\bf q}_{1},{\bf q}_{2};t)  \nonumber \\
f_{NE}\left( {\bf q}_{1},{\bf q}_{3};t\right)  &=&\exp \left[ -V_{NE}\left( 
{\bf q}_{1},{\bf q}_{2};t\right) \right] -1
\end{eqnarray}
and where the nonequilibrium potential $V_{NE}\left( {\bf q}_{1},{\bf q}%
_{2};t\right) $ is a sum of the equilibrium potential and terms related to
the moments of the velocity. This formulation has the desirable property
that, if linearized in the density by replacing $\ln y({\bf q}_{1},{\bf q}%
_{2};t)\simeq y({\bf q}_{1},{\bf q}_{2};t)-1$, it reduces to the PY
approximation in equilibrium. It is difficult, however, to see how to
incorporate the exact requirement of the information coming from the
collisional boundary condition and it is therefore an open question whether
this can give a model comparable to the nonequilibrium GMSA described above.
Indeed, in the derivation of this model, the authors explicitly neglect
velocity correlations of the kind used here to control the model of the
structure. Nevertheless, the solution of this model to allow for such a
comparison would be of some interest.

It is also appropriate to comment on the applicability of the Enskog-level
description of the system. The density chosen for the simulations is one at
which the Enskog description of elastic hard spheres is very good with most
quantities being accurately predicted to within a few percent. Furthermore,
neither the calculation of the pressure nor the pdf at contact requires
explicit knowledge of the one-body distribution at contact which is good,
because no exact solution to the Enskog equation for HCS exists. None the
less, the results described above show that there are systematic deviations
from the Enskog predictions at all values of the coefficient of restitution.
Evidence has also been given that these are at least partly due to the
presence of soft modes within the system. The conclusion is therefore mixed:
while the Enskog description seems to be qualitatively accurate, all but the
very smallest systems will contain soft or unstable modes that result in
significant deviations from it. This, more than the separation of time
scales discussed in ref.\cite{Tan98} would seem to be the greatest obstacle
to using a Boltzmann/Enskog description of the one body function or of using
a hydrodynamic description of the macroscopic state. Indeed, the Enskog and
hydrodynamic descriptions are successful in predicting fairly well the
location of the hydrodynamic instability in the vorticity. A better test of
these issues would be to study a related system, such as a sheared granular
fluid, which may be more stable.

The nonequilibrium GMSA described here works surprisingly well. Encouraging
results have also been found when this model was applied to simple sheared
fluids\cite{submitted2PRL} and a systematic study of this system is in
progress. It is also of interest to try to improve on the use of the
compressibility equation in the nonequilibrium model:\ one substitute would
be information coming from kinetic theory such as the asymptotic behavior of
the pdf for which various approaches are possible.

\appendix

\section{Origin of the collisional boundary condition}

\subsection{From BBGKY hierarchy}

It is instructive to derive the collisional boundary condition from the
BBGKY hierarchy, the first equation of which appears above as eq.(\ref
{2-body}). The second BBGKY equation is 
\begin{eqnarray}
&&\left[ \frac{\partial }{\partial t}+{\bf p}_{1}\cdot \frac{\partial }{%
\partial {\bf q}_{1}}+{\bf p}_{2}\cdot \frac{\partial }{\partial {\bf q}_{2}}%
-\overline{T}_{-}(x_{1},x_{2})\right] f^{(2)}(x_{1},x_{2};t) \\
&=&\int dx_{3}\;\left[ \overline{T}_{-}(x_{1},x_{3})+\overline{T}%
_{-}(x_{2},x_{3})\right] f^{(3)}(x_{1},x_{2},x_{3};t)
\end{eqnarray}
where the collision operator is 
\begin{eqnarray}
\overline{T}_{-}(x_{1},x_{2}) &=&\sigma _{0}^{2}\int_{\Omega }d\widehat{{\bf %
\sigma }}\;\Theta \left( {\bf p}_{12}\cdot \widehat{{\bf \sigma }}\right)
\;\left( {\bf p}_{12}\cdot \widehat{{\bf \sigma }}\right) \left[ \delta
\left( {\bf q}_{12}-\sigma _{0}\widehat{{\bf \sigma }}\right) \alpha ^{-2}%
\widehat{b}_{12}^{-1}-\delta \left( {\bf q}_{12}+\sigma _{0}\widehat{{\bf %
\sigma }}\right) \right]  \\
&=&\int d{\bf \sigma }\;\delta \left( \sigma -\sigma _{0}\right) \Theta
\left( {\bf p}_{12}\cdot \widehat{{\bf \sigma }}\right) \;\left( {\bf p}%
_{12}\cdot \widehat{{\bf \sigma }}\right) \left[ \delta \left( {\bf q}_{12}-%
{\bf \sigma }\right) \alpha ^{-2}\widehat{b}_{12}^{-1}-\delta \left( {\bf q}%
_{12}+{\bf \sigma }\right) \right]  \\
&=&\delta \left( q_{12}-\sigma _{0}\right) \left( {\bf p}_{12}\cdot {\bf q}%
_{12}\right) \left[ \Theta \left( {\bf p}_{12}\cdot {\bf q}_{12}\right)
\;\alpha ^{-2}\widehat{b}_{12}^{-1}+\Theta \left( -{\bf p}_{12}\cdot {\bf q}%
_{12}\right) \;\right]  \\
&=&\delta \left( q_{12}-\sigma _{0}\right) \left( {\bf p}_{12}\cdot {\bf q}%
_{12}\right) \left[ \alpha ^{-2}\widehat{b}_{12}^{-1}+1\;\right] \Theta
\left( -{\bf p}_{12}\cdot {\bf q}_{12}\right) 
\end{eqnarray}
We now observe that no matter what the state, the atoms cannot overlap so we
must be able to write the distribution as 
\begin{equation}
f^{(2)}(x_{1},x_{2};t)=\Theta \left( q_{12}-\sigma _{0}\right) \widetilde{f}%
^{(2)}(x_{1},x_{2};t)
\end{equation}
so that 
\begin{eqnarray}
&&\left[ {\bf p}_{1}\cdot \frac{\partial }{\partial {\bf q}_{1}}+{\bf p}%
_{2}\cdot \frac{\partial }{\partial {\bf q}_{2}}\right] \Theta \left(
q_{12}-\sigma _{0}\right) \widetilde{f}^{(2)}(x_{1},x_{2};t) \\
&=&{\bf p}_{12}\cdot {\bf q}_{12}\delta \left( q_{12}-\sigma _{0}\right) 
\widetilde{f}^{(2)}(x_{1},x_{2};t)+\Theta \left( q_{12}-\sigma _{0}\right) %
\left[ {\bf p}_{1}\cdot \frac{\partial }{\partial {\bf q}_{1}}+{\bf p}%
_{2}\cdot \frac{\partial }{\partial {\bf q}_{2}}\right] \widetilde{f}%
^{(2)}(x_{1},x_{2};t)
\end{eqnarray}
so that there are two sources of singularities in the second BBGKY equation:
one coming from the collision operator and one from the streaming operator.
If we integrate $q_{12}$ over a vanishingly small interval centered at $%
\sigma _{0}$, only the singular terms will contribute so that we conclude
they must cancel independently of the regular terms. This gives 
\begin{equation}
{\bf p}_{12}\cdot {\bf q}_{12}\delta \left( q_{12}-\sigma _{0}\right) 
\widetilde{f}^{(2)}(x_{1},x_{2};t)=\delta \left( q_{12}-\sigma _{0}\right)
\left( {\bf p}_{12}\cdot {\bf q}_{12}\right) \left[ \;\alpha ^{-2}\widehat{b}%
_{12}^{-1}+1\;\right] \Theta \left( -{\bf p}_{12}\cdot {\bf q}_{12}\right) 
\widetilde{f}^{(2)}(x_{1},x_{2};t)
\end{equation}
or 
\begin{equation}
\delta \left( q_{12}-\sigma _{0}\right) \Theta \left( {\bf p}_{12}\cdot {\bf %
q}_{12}\right) \widetilde{f}^{(2)}(x_{1},x_{2};t)=\delta \left(
q_{12}-\sigma _{0}\right) \;\alpha ^{-2}\widehat{b}_{12}^{-1}\Theta \left( -%
{\bf p}_{12}\cdot {\bf q}_{12}\right) \widetilde{f}^{(2)}(x_{1},x_{2};t)
\end{equation}
which is the desired result relating the post-collisional distribution, on
the left, to the pre-collisional distribution on the right. It is thus
apparent that this identity carries part of the information of the second
BBGKY hierarchy.

\subsection{From Conservation of probability}

It is clear that the probability to find two atoms moving towards each other
with a given relative momentum, the left hand side, is the same as that to
find two atoms in contact moving away from one another with the
corresponding post-collision momentum: 
\[
\delta \left( {\bf q}_{12}-\sigma \widehat{{\bf \sigma }}\right) \Theta
\left( -\widehat{{\bf q}}_{12}\cdot {\bf p}_{12}\right)
f_{2}(x_{1},x_{2},t)d^{3}x_{1}d^{3}x_{2}=\delta \left( {\bf q}_{12}-\sigma 
\widehat{{\bf \sigma }}\right) \Theta \left( \widehat{{\bf q}}_{12}\cdot 
{\bf p}_{12}^{\prime }\right) f_{2}(x_{1}^{\prime },x_{2}^{\prime
},t)d^{3}x_{1}^{\prime }d^{3}x_{2}^{\prime }
\]
which simply states that the probability to find two atoms in contact moving
towards each other with a given relative momentum, the left hand side, is
the same as that to find two atoms in contact moving away from one another
with the corresponding pre-collision momentum (the right hand side). Using 
\begin{eqnarray}
d^{3}x_{1}d^{3}x_{2} &=&\frac{1}{\alpha ^{2}}d^{3}x_{1}^{\prime
}d^{3}x_{2}^{\prime } \\
f_{2}(x_{1},x_{2},t) &=&\widehat{b}_{12}^{-1}f_{2}(x_{1}^{\prime
},x_{2}^{\prime },t)
\end{eqnarray}
this can be written as 
\begin{equation}
\delta \left( {\bf q}_{12}-\sigma \widehat{{\bf \sigma }}\right) \Theta
\left( \widehat{{\bf q}}_{12}\cdot {\bf p}_{12}^{\prime }\right) \frac{1}{%
\alpha ^{2}}\widehat{{\bf b}}_{12}^{-1}f_{2}(x_{1}^{\prime },x_{2}^{\prime
},t)=\delta \left( {\bf q}_{12}-\sigma \widehat{{\bf \sigma }}\right) \Theta
\left( \widehat{{\bf q}}_{12}\cdot {\bf p}_{12}^{\prime }\right)
f_{2}(x_{1}^{\prime },x_{2}^{\prime },t)
\end{equation}
. The factors of $\alpha $ in the Jacobian arise because of the change from
pre- to post-collisional momenta (${\bf p}_{12}^{\prime }=-\alpha {\bf p}%
_{12}$ gives one factor of $\alpha $) and because the relation between
positions ${\bf q}(t)$ after the collision and positions before the
collision also involves the momenta giving a second factor of $\alpha $.

\subsection{Evaluating correlations at contact}

From the identity and the assumption of molecular chaos, we have 
\begin{eqnarray}
\delta \left( {\bf q}_{12}-\sigma \widehat{{\bf \sigma }}\right)
Vf_{2}(x_{1},x_{2},t) &=&\delta \left( {\bf q}_{12}-\sigma \widehat{{\bf %
\sigma }}\right) f_{1}({\bf x}_{1};t)f_{1}({\bf x}_{2};t)g_{0}({\bf q}_{1},%
{\bf q}_{2};t) \\
&&+\delta \left( {\bf q}_{12}-\sigma \widehat{{\bf \sigma }}\right) \Theta
\left( \widehat{{\bf q}}_{12}\cdot {\bf p}_{12}\right) \left( \alpha ^{-2}%
\widehat{b}_{12}^{-1}-1\right) f_{1}(x_{1};t)f_{1}(x_{2};t)g_{0}({\bf q}_{1},%
{\bf q}_{2};t)
\end{eqnarray}
so that if averages over $f_{1}({\bf x}_{1};t)f_{1}({\bf x}_{2};t)g_{0}({\bf %
q}_{1},{\bf q}_{2};t)$ are denoted by $\left\langle ...\right\rangle _{0}$,
then for any two-body function $A=\sum_{j<i}A(x_{i},x_{j})\delta \left( {\bf %
q}_{ij}-\sigma \widehat{{\bf \sigma }}\right) $ one finds 
\begin{eqnarray}
A(\widehat{{\bf \sigma }}) &=&\left\langle A\right\rangle \\
&=&\left\langle A\right\rangle _{0}+\frac{N(N-1)}{2V}\int
dx_{1}dx_{2}\;\delta \left( {\bf q}_{12}-\sigma \widehat{{\bf \sigma }}%
\right) A(x_{1},x_{2})\Theta \left( \widehat{{\bf q}}_{12}\cdot {\bf p}%
_{12}\right) \left( \alpha ^{-2}\widehat{b}_{12}^{-1}-1\right)
f_{1}(x_{1};t)f_{1}(x_{2};t)g_{0}({\bf q}_{1},{\bf q}_{2};t) \\
&=&\left\langle A\right\rangle _{0}+\frac{N(N-1)}{2V}\int
dx_{1}dx_{2}\;\delta \left( {\bf q}_{12}-\sigma \widehat{{\bf \sigma }}%
\right) f_{1}(x_{1};t)f_{1}(x_{2};t)g_{0}({\bf q}_{1},{\bf q}_{2};t)\left(
\alpha ^{-1}\widehat{b}_{12}-1\right) \Theta \left( \widehat{{\bf q}}%
_{12}\cdot {\bf p}_{12}\right) A(x_{1},x_{2}) \\
&=&\left\langle A\right\rangle _{0}+\sum_{j<i}\left\langle \delta \left( 
{\bf q}_{ij}-\sigma \widehat{{\bf \sigma }}\right) \left( \Theta \left( -%
\widehat{{\bf q}}_{ij}\cdot {\bf p}_{ij}\right) \alpha ^{-1}\widehat{b}%
_{ij}-\Theta \left( \widehat{{\bf q}}_{ij}\cdot {\bf p}_{ij}\right) \right)
A(x_{i},x_{j})\right\rangle _{0}
\end{eqnarray}
and, in particular, if $A(x_{i},x_{j})=\frac{1}{4\pi n^{2}}$ then 
\begin{eqnarray}
g(\widehat{{\bf \sigma }}) &=&g_{0}(\widehat{{\bf \sigma }})+\frac{1-\alpha 
}{2\alpha }g_{0}(\widehat{{\bf \sigma }}) \\
&=&\frac{1+\alpha }{2\alpha }g_{0}(\widehat{{\bf \sigma }})
\end{eqnarray}
as reported in the text.

\end{document}